\documentclass[a4paper,11pt]{article}

\usepackage{pos}

\usepackage{physics}
\usepackage{tensor}
\usepackage{hyperref}

\DeclareMathAlphabet{\mathbb}{U}{msb}{m}{n}

\renewcommand{\logo}{\relax}

\newcommand{\rn}{\mathbb{R}}

\newcommand{\df}{\coloneqq}

\newcommand{\SOP}{\text{SO}(1,3)^\uparrow}

\newcommand{\td}[1]{\tilde{#1}}

\newcommand{\lag}{\mathscr{L}}

\newcommand{\heta}{\hat{\eta}}

\newcommand{\hps}{\hat{\psi}}

\newcommand{\hph}{\hat{\phi}}

\newcommand{\hpm}{\hat{P}}

\newcommand{\hJ}{\hat{J}}

\newcommand{\hpr}{\hat{\mathcal{P}}}

\newcommand{\thps}[2]{\hat{\td{\psi}}^{#1} (#2)}

\newcommand{\thph}[3]{\hat{\td{\phi}}^{#1}_{#2} (#3)}

\newcommand{\hh}{\hat{H}}

\newcommand{\hlag}{\hat{\lag}}

\newcommand{\vx}{\vec{x}}

\newcommand{\der}[3]{\partial_{#1}^{#2}{#3}}

\newcommand{\be}{\begin{equation}}

\newcommand{\ee}{\end{equation}}

\newcommand{\bea}{\begin{eqnarray}}

\newcommand{\eea}{\end{eqnarray}}

\title{Flavour oscillations in pseudo-Hermitian\\ quantum theories}

\author[a]{Robert Mason}
\author*[b]{Peter Millington}
\author[b]{Esra Sablevice}

\affiliation[a]{Department of Mathematical Sciences, University of Liverpool,\\ Liverpool L69 7ZL, United Kingdom}

\affiliation[b]{Department of Physics and Astronomy, University of Manchester,\\Manchester M13 9PL, United Kingdom}

\emailAdd{robert.mason@liverpool.ac.uk}
\emailAdd{peter.millington@manchester.ac.uk}
\emailAdd{esra.sablevice@manchester.ac.uk}

\abstract{
This note summarises recent progress in the formulation of flavour mixing and oscillations in pseudo-Hermitian quantum theories with non-Hermitian mass mixing matrices. Such non-Hermitian quantum theories are made viable by the existence of a discrete anti-linear symmetry of the Hamiltonian, which ensures that states have real energies. We describe oscillation and survival probabilities in a non-Hermitian two-state quantum mechanical system that are consistent with unitarity, and highlight features of these pseudo-Hermitian flavour oscillations that are unique compared to their Hermitian counterparts.
}

\FullConference{The European Physical Society Conference on High Energy Physics (EPS-HEP2023)\\
 21-25 August 2023\\
Hamburg, Germany\\}

\renewcommand{\hookAfterAbstract}{%
\par\bigskip
LTH 1355
}

\begin{document}
\maketitle


\section{Introduction}

In ``standard'' quantum mechanics, the Hamiltonian operator $\hat{H}$ is assumed to be Hermitian, i.e., $\hat{H}^{\dag}=\hat{H}$, where $\dag$ is the composition of complex conjugation and transposition. This guarantees real energies and unitary time evolution. However, it is known that real energies and unitary time evolution can be guaranteed by the weaker condition of pseudo-Hermiticity~\cite{Mostafazadeh:2001jk, Mostafazadeh:2001nr, Mostafazadeh:2002id}. The viability of a pseudo-Hermitian quantum theory relies on the existence of a Hermitian operator $\heta$, such that $\hat{H}^{\dag}=\hat{\eta}\hat{H}\hat{\eta}^{-1}$. The Hamiltonian is then said to be $\heta$-pseudo-Hermitian. The inner product $\braket{\cdot}{\cdot}_{\hat{\eta}}\coloneq \braket{\cdot} {\heta \cdot}$  yields real expectation values, even if the eigenenergies are complex. A pseudo-Hermitian quantum theory has three regimes:\ (i) unbroken anti-linear symmetry, with real eigenenergies and orthogonal eigenvectors w.r.t.\ $\braket{\cdot}{\cdot}_{\hat{\eta}}$; (ii) broken anti-linear symmetry, in which the spectrum contains complex-conjugate pairs of eigenvalues, whose eigenvectors have vanishing norm; and (iii) exceptional points, where the Hamiltonian becomes defective. 

Pseudo-Hermitian quantum mechanics~\cite{Mostafazadeh:2001jk, Mostafazadeh:2001nr, Mostafazadeh:2002id} has found applications in many areas of physics, from optics through to condensed matter physics (for reviews, see, e.g., Refs.~\cite{ElGanainy, Ashida}). This is particularly true of the special case of $\mathcal{PT}$-symmetric quantum mechanics~\cite{Bender:2005tb}, in which the viability of the quantum theory is ensured by the symmetry of the Hamiltonian under the combined action of parity $\mathcal{P}$ and time-reversal $\mathcal{T}$, with the latter being represented by an anti-linear operator.

The area of pseudo-Hermitian quantum field theory (pseudo-Hermitian QFT), however, is less developed, and this forms the primary focus of this note. We describe the consistent, first-principles formulation of a pseudo-Hermitian QFT of two complex scalar fields with a non-Hermitian mass mixing in Sec.~\ref{sec:formulation}, based on Ref.~\cite{Sablevice:2023odu}.  We then compare the transition probabilities between the two flavour states of this system to those of a Hermitian model with two-state mixing in Sec.~\ref{sec:oscillations}, based on Ref.~\cite{Alexandre:2023afi}. Flavour oscillations in non-Hermitian systems have attracted particular attention in the context of neutrino physics~\cite{Jones-Smith:2009qeu, Alexandre:2015kra}.


\section{Pseudo-Hermitian quantum field theory}
\label{sec:formulation}

We start by considering the Heisenberg equation of motion for an $n$-component quantum field operator $\hat{\psi}^a$, and the Hermitian-conjugate equation:
\be
    \left[\hat{\psi}^a(\vx,t),\hat{H}\right]=i\partial_t\hat{\psi}^a(\vx,t)\quad \Leftrightarrow \quad   \left[\hat{\psi}^{\dag a}(\vx,t),\hat{H}^{\dag}\right]=i\partial_t\hat{\psi}^{\dag a}(\vx,t)\;.
\ee
We see that the `conjugate' field operator $\hat{\psi}^{\dag a}$ evolves with the Hermitian conjugate $\hh^{\dag}$ of the Hamiltonian $\hh$. Hence, if the Hamiltonian is non-Hermitian $\hh^{\dag}\neq \hh$, the time evolutions of $\hat{\psi}^a$ and $\hat{\psi}^{\dag a}$ are not governed by the same Hamiltonian (see Ref.~\cite{Alexandre:2020gah}), and an action formulated from the pair  $(\hat{\psi}^a,\hat{\psi}^{\dag a})$ will not lead to a consistent, canonical formulation of Hamiltonian and Lagrangian dynamics (see Ref.~\cite{Alexandre:2017foi}). Moreover, the theory will not be invariant under the symmetry group of QFT in Minkowski spacetime:\ the proper Poincar\'e group.

The proper Poincar\'e group I$\SOP=\SOP\rtimes \rn^{1,3}$ consists of proper Lorentz transformations $\SOP$, with six generators $\hJ^{\mu\nu}$, and spacetime translations $\rn^{1,3}$, with four generators $\hpm^{\mu}$, where the Hamiltonian $\hh=\hpm^{0}$ is the generator of time translations. These generators are not independent, but are related through the Poincar\'e algebra. For example, if we conjugate one of the Lie brackets, e.g.,
\be
 [\hh,\hJ^{0 i}]=i\hpm^{i}\implies
 [\heta\hh\heta^{-1},\hJ^{0 i \dag}]=i\hpm^{i\dag}\implies  [\hh,\heta^{-1}\hJ^{0 i \dag}\heta]=i\heta^{-1}\hpm^{i\dag}\heta\;,
\ee
we see that non-Hermiticity of the Hamiltonian impacts non-Hermiticity of the remaining group generators. In fact, they must all be $\heta$-pseudo-Hermitian, not just the Hamiltonian~\cite{Sablevice:2023odu}:
\be
\label{eq:non_Hermiticity_of_other_generators}
\hJ^{\mu\nu \dag}=\heta\hJ^{\mu\nu}\heta^{-1},\:\:\:\hpm^{\mu\dag}=\heta\hpm^{\mu}\heta^{-1}\;.
\ee
Note that the generators remain Hermitian if they commute with $\heta$. Thus, $\hps^{a}$ and $\hps^{\dag a}$ transform under a different set of representations of the proper Poincar\'e group~\cite{Sablevice:2023odu}:
\begin{subequations}
\begin{align}
 [\hps^{a} (\vx,t),\hat{P}_{\mu}]=& \:i\partial_{\mu}\hps^{a} (\vx,t)\:,\:\:\:\:\:\:\:\:\:\:\:
 [\hps^{a} (\vx,t),\hJ^{\mu\nu}]= \:(M^{\mu\nu})\indices{^a_b}\hps^{b} (\vx,t)\;,\\
  [\hat{\psi}^{\dag a} (\vx,t),\hpm_{\mu}^ {\dag}]=&\:i\der{\mu}{}{\hat{\psi}^{\dag a} (\vx,t)}\;,\:\:\:\:\:\:
 [\hps^{\dag a} (\vx,t),\hJ^{\dag \mu\nu}]=\hps^{\dag b} (\vx,t) (- M^{\dag \mu \nu})\indices{_b^a}\;,
\end{align}
\end{subequations}
where $M^{\mu\nu}$ is the $n\times n$ matrix representation of the proper Lorentz group.

A consistent formulation of a non-Hermitian QFT must be composed of quantum fields transforming under the same representation of the proper Poincar\'e group~\cite{Chernodub:2021waz}. Thus, following Ref.~\cite{Sablevice:2023odu}, we aim to find a `dual' quantum field $\hat{\td{\psi}}^{\dag}$, which transforms in the same representation as $\hps$:
\be
[\thps{\dag a}{\vx,t},\hpm_{\mu}]=i\partial_{\mu}\thps{\dag a}{\vx,t}\:,\:\:\:[\thps{\dag a}{\vx,t},\hJ^{\mu\nu}]=\thps{\dag b}{\vx,t}(-M^{\mu\nu})\indices{_b^a}\;.
\ee
Assuming the generators of an $n$-dimensional matrix representation are $\pi$-pseudo-Hermitian w.r.t.\ some Hermitian $n\times n$ matrix $\pi$, i.e., $M^{\mu\nu\dag}=\pi M^{\mu\nu} \pi^{-1}$, 
the dual quantum field is of the form~\cite{Sablevice:2023odu}
\be
\label{eq:dual_field}
\hat{\tilde{\psi}}^{\dag a}(x)\df\heta^{-1}\hat{\psi}^{\dag b} (x_{\eta})\heta \:\pi\indices{_b^a}\;.
\ee
Here, $x_{\eta}$ is the transformed coordinate w.r.t.\ the operator $\heta$, which might, e.g., be parity $x_{P}$. The definition of the dual field in Eq.~\eqref{eq:dual_field} is valid for fields of any spin~\cite{Sablevice:2023odu}.

We illustrate this formulation using a simple example with Lagrangian density (cf.~Ref.~\cite{Alexandre:2017foi}):
\be
\hlag=\partial_{\mu}\thph{\dag}{}{x}\partial^{\mu}\hph (x)-\thph{\dag}{}{x}M^{2}\hph (x)\;,\:\:\:
M^{2}=\mqty(m_{1}^{2} & m_5^{2}\\ -m_5^{2} & m^{2}_{2})\:,\:\:\:P=\mqty(1 & 0\\0&-1)\;,\label{eq:scalarLag}
\ee
where $m_1^2>m_2^2>0$ and $m_5^2>0$. The operator $\hph=(\hph_{1},\hph_{2})$ is a two-component complex scalar field composed of a scalar $\hph_{1}$ and a pseudo-scalar $\hph_{2}$. The squared mass matrix $M^{2}\neq M^{2\dag}$ is $P$-pseudo-Hermitian w.r.t.\ the parity matrix $P$, i.e., $M^{2\dag}=PM^{2}P^{-1}$, and its eigenvalues
\be
    m_{\pm}^2=(m_1^2+m_2^2)/2\pm\left[(m_1^2-m_2^2)^2/4-m_5^4\right]^{1/2}
\ee
are real when the argument of the square root is non-negative. 

The classical Lagrangian corresponding to Eq.~\eqref{eq:scalarLag} is $\mathcal{PT}$-symmetric, while the quantum Lagrangian (and Hamiltonian) is $\hpr$-pseudo-Hermitian. The parity operator is an indefinite metric operator that yields negative probability norms. However, in the regime where the squared mass eigenvalues are real, we can construct a matrix $A$
that commutes with the squared mass matrix $[M^{2},A]=0$, and $M^{2}$ is $PA$-pseudo-Hermitian.\footnote{We use the notation of Ref.~\cite{Chernodub:2021waz}; the corresponding transformations are referred to as $\mathcal{C}$~\cite{Bender:2005tb, Sablevice:2023odu} or $\mathcal{C}'$~\cite{Alexandre:2023afi, Alexandre:2020gah} elsewhere.} This matrix is~\cite{Alexandre:2020gah}
\be
    A=\frac{1}{\sqrt{1-\zeta^2}}\begin{pmatrix} 1 & \zeta \\ -\zeta & -1\end{pmatrix}\quad \text{with} \quad  A^2=\mathbb{I}\;,\quad PA=\frac{1}{\sqrt{1-\zeta^2}}\begin{pmatrix} 1 & \zeta \\ \zeta & 1\end{pmatrix}\quad \text{and} \quad \zeta= \frac{2m_5^2}{m_1^2-m_2^2}\;.
\ee
Similarly, we find another operator $\hat{\mathcal{A}}$ that commutes with the Hamiltonian $[\hh,\hat{\mathcal{A}}]=0$, and the Lagrangian (and Hamiltonian) is $\hpr \hat{\mathcal{A}}$-pseudo-Hermitian. This gives a natural choice for the operator $\heta=\hpr\hat{\mathcal{A}}$ and the matrix $\pi=PA$,\footnote{This should be compared with the formulation in Ref.~\cite{Alexandre:2020gah} in which the dual field is defined w.r.t.\ parity.} so that the dual field is~\cite{Sablevice:2023odu}
\be
\thph{\dag a}{}{x}=(\hpr\hat{\mathcal{A}})\hat{\phi}^{\dag b}(x) (\hpr \hat{\mathcal{A}})^{-1} (PA)\indices{_b^a}\quad \text{with}\quad x_{PA}=x\;.
\ee
Since the Hamiltonian is $\hpr\hat{\mathcal{A}}$-pseudo-Hermitian and $\hpr\hat{\mathcal{A}}$ is a positive-definite metric operator, energy eigenstates are orthonormal w.r.t.\ the inner product
\be
\langle \cdot \vert \cdot\rangle_{\hpr\hat{\mathcal{A}}}=\langle \cdot \vert \hpr\hat{\mathcal{A}}\:\cdot\rangle \;.
\ee


\section{Oscillation probabilities}
\label{sec:oscillations}

We now turn to transition probabilities in the two-flavour system in Eq.~\eqref{eq:scalarLag}, following Ref.~\cite{Alexandre:2023afi}. For simplicity, we consider only the zero-momentum modes.

The squared mass matrix in Eq.~\eqref{eq:scalarLag} is diagonalised via~\cite{Chernodub:2021waz}
\be
    S^{-1}M^2S={\rm diag}(m_+^2,m_-^2)\quad \text{where}\quad S=\begin{pmatrix} \cosh(\theta) & -\sinh(\theta) \\ -\sinh(\theta) & \cosh(\theta)\end{pmatrix}\;,
\ee
with $\theta=\mathrm{arctanh}(\zeta)/2$. It is then tempting to take the ``flavour'' states to be~\cite{Alexandre:2023afi}
\be
\label{eq:flavourstates}
\ket{\phi_{1(2)}(t)}=\cosh(\theta)\,e^{im_{+(-)}t}\ket{\phi_+(0)}+\sinh(\theta)\,e^{im_{-(+)}t}\ket{\phi_-(0)} \; ,
\ee
where $\ket{\phi_{+(-)}(0)}$ are the eigenstates of the squared mass matrix given by~\cite{Alexandre:2017foi}
\be
\label{e1e2}
\ket{\phi_+(0)}=N\begin{pmatrix} \zeta\;, & -1+\sqrt{1-\zeta^2} \end{pmatrix}\quad\text{and}\quad \ket{\phi_-(0)}=N\begin{pmatrix} -1+\sqrt{1-\zeta^2}\;, & \zeta \end{pmatrix}\;,
\ee
and $N$ is a normalisation constant. At $t=0$, $\ket{\phi_{1}(0)} \propto (1,0)$ and $\ket{\phi_{2}(0)} \propto (0,1)$, as we would expect. However, the oscillation (e.g., $1\to 2$) and survival (e.g., $1\to 1$) probabilities given by
\be
    \mathbb{P}_{1\to i}(t,t_0) = 
    \langle \phi_i(t) \vert \phi_1(t_0) \rangle_{\hpr\hat{\mathcal{A}}}\langle \phi_1(t_0) \vert \phi_i(t) \rangle_{\hpr\hat{\mathcal{A}}}
\ee
can be negative or larger than unity (cf.~Refs.~\cite{Ohlsson:2019noy, Alexandre:2020gah, Alexandre:2023afi}). The reason for this is that the states $\ket{\phi_{1}(t)}$ and $\ket{\phi_{2}(t)}$ do not form an orthonormal basis with respect to $\hpr\hat{\mathcal{A}}$.

Instead, an orthonormal basis w.r.t.\ the $\hpr\hat{\mathcal{A}}$ inner product is obtained by spanning the flavour space (of zero-momentum states) with $\{\ket{\phi^{\phantom{\mathcal{A}\!\!\!}}_1},\ket{\phi_2^\mathcal{A}}\}$ or $\{\ket{\phi_1^\mathcal{A}},\ket{\phi_2^{\phantom{\mathcal{A}\!\!\!}}}\}$~\cite{Alexandre:2023afi}. Such a basis is expected to be selected by any $\hpr \hat{\mathcal{A}}$-pseudo-Hermitian interaction terms appended to the Lagrangian density in Eq.~\eqref{eq:scalarLag}. Proceeding in this way, survival probabilities are calculated w.r.t.\ the $\hpr\hat{\mathcal{A}}$ inner product, but oscillation probabilities are effectively calculated using the states in Eq.~\eqref{eq:flavourstates} w.r.t.\ the $\hpr$ inner product. We then find positive and unitary probabilities in the $\mathcal{PT}$-unbroken regime ($|\zeta|\leq 1$)~\cite{Alexandre:2023afi}:
\be
\mathbb{P}_{1\to2}(t,t_0)=\langle \phi_2^{\mathcal{A}}(t) \vert \phi_1(t_0) \rangle_{\hpr\hat{\mathcal{A}}}\langle \phi_1(t_0) \vert \phi_2^{\mathcal{A}}(t) \rangle_{\hpr\hat{\mathcal{A}}}=\zeta^2\sin^2\left[\Delta \omega \Delta t/2\right]\geq 0\;, 
\ee
with $\mathbb{P}_{1\to1}(t,t_0)=1-\mathbb{P}_{1\to2}(t,t_0)$, where $\Delta \omega = m_+ - m_-$, $\Delta t = t-t_0$, and $N=[2(1-\sqrt{1-\zeta^2})]^{-1/2}$ has been chosen such that $\mathbb{P}_{1\to1}(t_0,t_0)=1$. Here, $\bra{\phi_i}=(\ket{\phi_i})^{\dag}$, cf.~Ref.~\cite{Alexandre:2020gah}.

\begin{figure}
\centering
\includegraphics[width=0.75\textwidth,trim= 0 3.5cm 0 3.5cm ,clip]{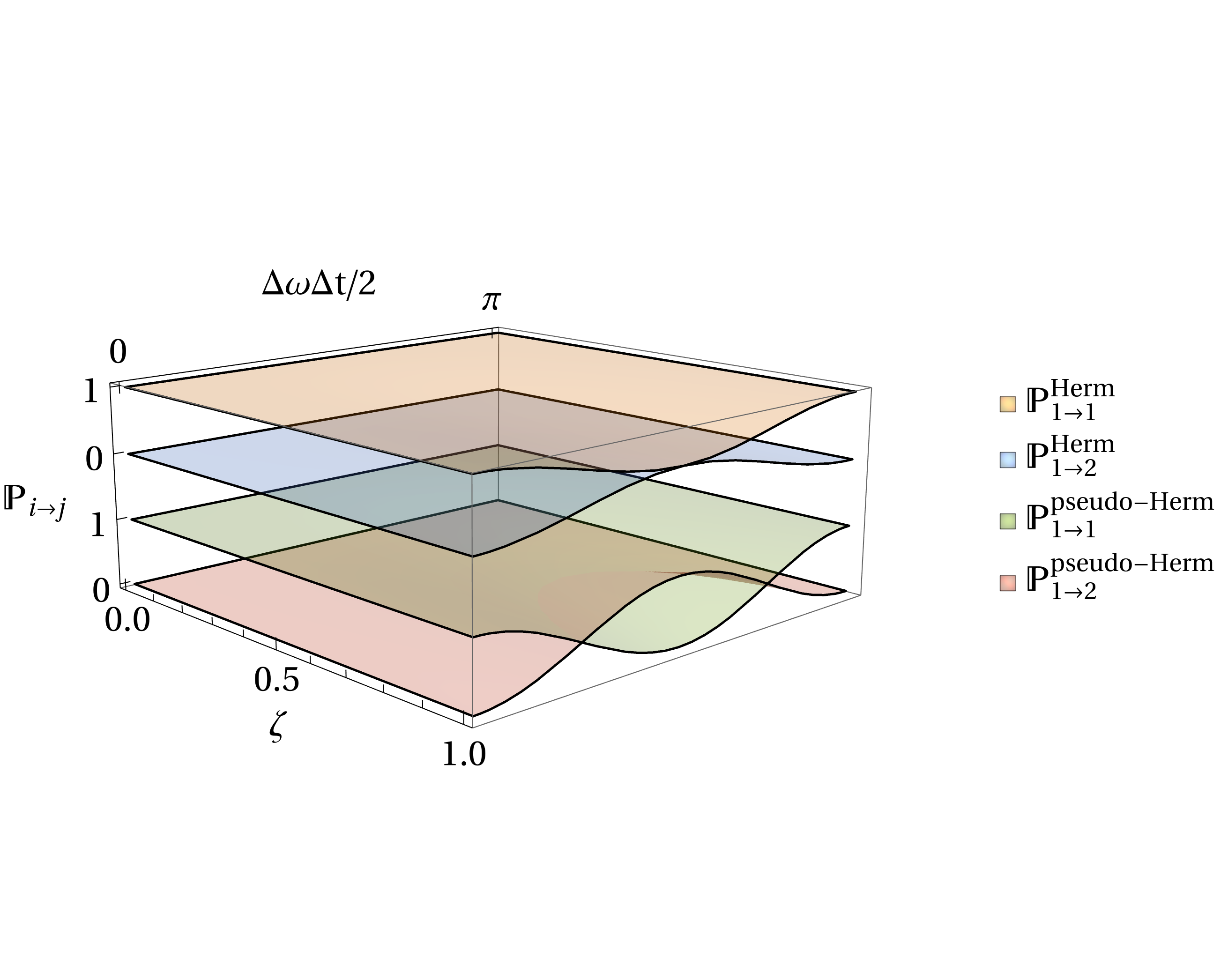} 
\caption{Oscillation and survival probabilities for the pseudo-Hermitian two-state model ($\mathbb{P}^{\text{pseudo-Herm}}_{1\to i}$) and the  Hermitian analogue ($\mathbb{P}_{1\to i}^{\text{Herm}}$), as a function of $\Delta\omega \Delta t /2$ and $\zeta$, based on Ref.~\cite{Alexandre:2023afi}.}
\label{fig:1}
\end{figure}

In Fig.~\ref{fig:1}, and as per Ref.~\cite{Alexandre:2023afi}, we compare the oscillation and survival probabilities of the pseudo-Hermitian model in Eq.~\eqref{eq:scalarLag} with a Hermitian model with squared mass matrix
\bea
\label{eq:HermMass}
    M^2_{\rm Herm}\ =\ \begin{pmatrix} m_1^2 & m_5^2 \\ m_5^2 & m_2^2\end{pmatrix} \quad \text{for which} \quad \mathbb{P}_{1\to 2}^{\text{Herm}} = \frac{\zeta^2}{1+\zeta^2}\sin^2\left[\Delta \omega \Delta t/2\right]\;.
\eea
For the pseudo-Hermitian case, the transition probabilities can reach unity for $\zeta\to\pm 1$ or $m_+^2=m_-^2$. This occurs for the  transition probabilities of the Hermitian model (calculated using the usual Hermitian inner product) only when $\zeta \to \pm \infty$ or \smash{$m_{+,\,{\rm Herm}}^2/(m_1^2+m_2^2)\to \infty$}, \smash{$m_{-,\,{\rm Herm}}^2/(m_1^2+m_2^2)<0$}, where \smash{$m_{\pm,\,{\rm Herm}}^2=(m_1^2+m_2^2)/2\pm\left[(m_1^2-m_2^2)^2/4+m_5^4\right]^{1/2}$}.

\section{Concluding remarks}

We have outlined the formulation of a pseudo-Hermitian quantum field theory of two complex scalars with non-Hermitian flavour mixing, and described the transition probabilities of the corresponding two-state quantum-mechanical system. These probabilities are consistent with time-translational invariance and unitarity. Interestingly, they exhibit a very different relationship between Lagrangian parameters and physical observables compared with an ab initio Hermitian theory, in this case in the relationship between the eigenmasses and the effective mixing angle. These results evidence the potential phenomenological relevance of pseudo-Hermitian quantum field theories.


\acknowledgments

This work was supported by an EPSRC PhD studentship [Grant No.~EP/R513271/1] (RM), by the Science and Technology Facilities Council (STFC) [Grant No.~ST/X00077X/1] (PM), a United Kingdom Research and Innovation (UKRI) Future Leaders Fellowship [Grant No.~MR/V021974/2] (PM), and the University of Manchester (ES). RM and PM thank Jean Alexandre, Madeleine Dale and John Ellis for their collaboration on Ref.~\cite{Alexandre:2023afi} outlined herein.
The authors thank Jean Alexandre for comments on the manuscript.
For the purpose of open access, the authors have applied a Creative Commons Attribution (CC BY) licence to any Author Accepted Manuscript version arising.


\section*{Data Access Statement}

No data were created or analysed in this study.


\end{document}